\documentclass[11pt]{llncs}

\usepackage{amssymb}
\usepackage{amsmath}
\usepackage{url}
\usepackage{array}
\usepackage{multirow}
\usepackage{graphicx}
\usepackage{times}
\usepackage{fullpage}

\newcommand{\ket}[1]{|{#1}\rangle}
\newcommand{\bra}[1]{\langle{#1}|}

\newcommand{\ssp}{\vspace*{1mm}}
\newcommand{\bsp}{\vspace*{2mm}}

\begin{document}
\pagestyle{plain}
\title{Applying Grover's algorithm to AES:\newline quantum resource estimates}

\author{Markus Grassl$^1$\and Brandon Langenberg$^2$
\and Martin Roetteler$^3$\and Rainer~Steinwandt$^{2}$}
\institute{Universit\"at Erlangen-N\"urnberg \& Max Planck Institute for the Science of Light, \\G\"unther-Scharowsky-Stra{\ss}e 1, Bau~24, 91058 Erlangen, Germany, \email{Markus.Grassl@fau.de}
\and
Florida Atlantic University, 777~Glades~Road, Boca Raton, FL 33431, U.S.A., \email{\{blangenb,rsteinwa\}@fau.edu}\and
Microsoft Research, One Microsoft Way, Redmond, WA 98052, U.S.A., \email{martinro@microsoft.com}
}

\maketitle
\begin{abstract} We present quantum circuits to implement an exhaustive key search for the Advanced Encryption Standard (AES) and analyze the quantum resources required to carry out such an attack. We consider the overall circuit size, the number of qubits, and the circuit depth as measures for the cost of the presented quantum algorithms. Throughout, we focus on Clifford$+T$ gates as the underlying fault-tolerant logical quantum gate set. In particular, for all three variants of AES (key size 128, 192, and 256~bit) that are standardized in FIPS-PUB 197, we establish precise bounds for the number of qubits and the number of elementary logical quantum gates that are needed to implement Grover's 	quantum algorithm to extract the key from a small number of AES plaintext-ciphertext pairs.  \ssp
 
\end{abstract}

\section{Introduction} 
Cryptanalysis is an important area where quantum algorithms have found applications. Shor's seminal work invalidates some well-established computational assumptions in \emph{asymmetric} cryptography \cite{Sho97}, including the hardness of factoring and the computation of discrete logarithms in finite cyclic groups such as the multiplicative group of a finite field. On the other hand, regarding \emph{symmetric} encryption, the impact of quantum algorithms seems less dramatic. While a quantum version of related key attacks \cite{RoSt15} would be a threat for block ciphers provided that quantum access to the encryption function is given, as this requires the ability to generate quantum superpositions of related keys, this attack model is somewhat restrictive. In particular, the related key attack of \cite{RoSt15} is not applicable to, say, a context where a small number of plaintext-ciphertext pairs are given and the goal is to identify the encryption key. 

It has been known for some time that in principle Grover's search algorithm \cite{Gro96} can be applied to the problem of finding the key: the square root speed-up offered by Grover's algorithm over a classical exhaustive key search seems to be the most relevant quantum cryptanalytic impact for the study of block ciphers. To actually implement such an attack, the Boolean predicate that is queried in Grover's algorithm needs to be realized as a circuit. 
Perhaps interestingly, even for the most obvious target---the Advanced Encryption Standard \cite{FIPS01}, which in its 256-bit version has recently been suggested to be quantum-safe \cite{Auetal15}---to the best of our knowledge no detailed logical level resource estimate for implementing Grover's algorithm is available. The seemingly simple task of implementing the AES function actually requires some analysis as the circuit implementation is required to be reversible, i.e., it must be possible to implement the operation via an embedding into a permutation. Once a reversible implementation is known, in principle also a quantum implementation can be derived as the set of permutations is a subset of all unitary operations. 

\paragraph{Our contribution.} We provide reversible circuits that implement the full Advanced Encryption Standard AES-$k$ for each standardized key size (i.e., $k=128, 192, 256$). We establish resource estimates for the number of qubits and the number of Toffoli gates, controlled NOT gates, and NOT gates. See \cite{NC:2000} for basic definitions of quantum and reversible logic gates. 
Furthermore, we consider decompositions of the reversible circuits into a universal fault-tolerant gate set that can then be implemented as the set of logical gates. As underlying fault-tolerant gate set we consider the so-called set of Clifford$+T$ gates.\footnote{As is common, we do not distinguish between $T=\left(\begin{array}{cc}1&0\\0&\exp(i\pi/4)\end{array}\right)$ and $T^\dagger$-gates.} This gate set is motivated, e.g., by the fact that this set of gates can be implemented fault-tolerantly on a large set of codes, including the surface code family \cite{FSG:2009,Fowl12f} and concatenated CSS codes \cite{Stea03b,Reichardt:2009}. Clifford gates typically are much cheaper than the $T$-gate which commonly is implemented using state distillation. When breaking down the circuit to the level of $T$-gates we therefore pay attention to reducing the overall $T$-count. See also \cite{AMMR13,AMMR13b} for techniques how to optimize the $T$-count and \cite{AMM:2014} for techniques that allow to navigate the trade-space between $T$-depth and the number of qubits used. 
For the particular case of the Toffoli gate we use an implementation that requires $7$ $T$-gates and several Clifford gates, see \cite{NC:2000,AMMR13}. There is a probabilistic circuit known that implements the Toffoli gate with only $4$ $T$-gates \cite{JonesToff}, however, as the architecture requirements will be stronger in that measurement and feed-forward of classical information is required, we focus on the purely unitary decomposition that requires $7$ $T$-gates. We remark however, that the only source of $T$-gates in this paper are Toffoli gates, hence it is possible to use Jones' Toffoli factorization {\em mutatis mutandis} which leads to all given resource estimates for the $T$-count being multiplied by $4/7$ and the requirement of $1$ additional ancilla qubit.
In our resource estimates we do not to restrict interactions between qubits and leave the implementation, e.g., on a 2D nearest neighbor array for further study, including an investigation of the remaining \emph{quantum circuit placement} problems \cite{MFM07} that will have to be solved for the logical gate lists that are produced by our approach. 

One of our main findings is that the number of logical qubits required to implement a Grover attack on AES is relatively low, namely between around $3,000$ and $7,000$ logical qubits. However, due to the large circuit depth of unrolling the entire Grover iteration, it seems challenging to implement this algorithm on an actual physical quantum computer, even if the gates are not error corrected. It is worth noting that much of the circuit cost within each Grover iteration originates from the key expansion, i.\,e., from deriving the round keys and that the overall depth is a direct result of the serial nature of Grover's algorithm. 

\section{Preliminaries: Grover's algorithm} 
Before going into technicalities of how to implement AES as a quantum
circuit, we briefly recall the interface that we need to provide to
realize a key search, namely Grover's algorithm \cite{Gro96}. The
Grover procedure takes as an input a quantum circuit implementing a
Boolean function $f\colon\{0,1\}^k \longrightarrow\{0,1\}$ in the usual way, i.e.,
via a quantum circuit $U_f$ that implements $\ket{x}\ket{y} \mapsto
\ket{x}\ket{y \oplus f(x)}$, where $x\in \{0,1\}^n$ and $y \in
\{0,1\}$. The basic Grover algorithm finds an element $x_0$ such that
$f(x_0)=1$. Denoting by $H$ the $2\times 2$ Hadamard
transform, the Grover algorithm consists of repeatedly applying the
operation $G$ to the initial state $\ket{\psi}\otimes \ket{\varphi}$,
where $\ket{\psi} = \frac{1}{\sqrt{2^k}} \sum_{x\in \{0,1\}^k}
\ket{x}$, $\ket{\varphi} = \frac{1}{\sqrt{2}}(\ket{0}-\ket{1})$, and
where $G$ is defined as
\begin{equation}\label{eq:grover}
G = U_f \, \left((H^{\otimes k} (2 \ket{0}\bra{0}-{\mathbf 1}_{2^k}) H^{\otimes k}) \otimes {\mathbf 1}_2\right),
\end{equation}
where $\ket{0}$ denotes the all zero basis state of the appropriate
size.  Overall, $G$ has to be applied a number of $O(\sqrt{N/M})$
times in order to measure an element $x_0$ such that $f(x_0)=1$ with
constant probability, where $N$ is the total number of candidates,
i.e., $N=2^k$, and provided that there are precisely $M$ solutions, i.e.,
$M = |\{x\colon f(x)=1\}|$; see also \cite[Section~6.1.2]{NC:2000}, \cite{BBH+:98}
for an analysis. If we know that there is only one solution, i.e.,
$M=1$, this means that we can find a solution by applying $H^{\otimes
  k+1}$ to the initial state $\ket{0}^{\otimes k}\otimes \ket{1}$ and
then applying $G^{\ell}$, where $\ell = \lfloor \frac{\pi}{4} \sqrt{N}
\rfloor$, followed by a measurement of the entire quantum register
which will yield a solution $x_0$ with high probability
\cite[Section~6.1.4]{NC:2000}, \cite{BBH+:98}.

As we will show in the following section, we can indeed define a
function $f$ from the set of possible keys, i.e., $k\in\{128,192,256\}$ for the case
of AES, such that there is (plausibly) precisely one solution to the problem of
finding the correct key $K$ that was used to encrypt a small set of
given plaintext-ciphertext pairs, i.e., we can (plausibly) enforce the situation
$M=1$ by defining a suitable function $f$. We remark, however, that it
is possible to modify Grover's algorithm in various ways so that it
can cope with a larger (but known) number $M>1$ of solutions or even
with a completely unknown number of solutions: as mentioned above, if
the number $M$ of solutions is known, $O(\sqrt{N/M})$ iterations are
enough, however, if the number is {\em unknown}, there is an issue
that it is not possible to pick the right number of iterations a
priori. Nonetheless, there is a variant of the algorithm which finds a
solution in expected running time $O(\sqrt{N/M})$ even when the number
$M$ of solutions is unknown \cite[Section 6]{BBH+:98}.

There are several ways out of this dilemma which we mention
briefly for completeness but point out that we did not
implement these alternatives: one can first apply a quantum algorithm
to count the number of solutions \cite{GHT:98,BBH+:98} or one can do
an exponential search on the number of iterations
\cite{BBH+:98,BHM+:2002}, or one can employ an adaptive schedule in
which the Grover operator is changed to an operator that rotates by
different angles depending on the index of the iteration
\cite{YLC:2014}, thereby driving the oscillation of the quantum state
into a bounded region (the ``fixed point'') which then yields a
solution upon measurement.

\begin{figure}
\centerline{
\begin{tabular}{c@{\qquad}c}
\includegraphics[height=1.1in]{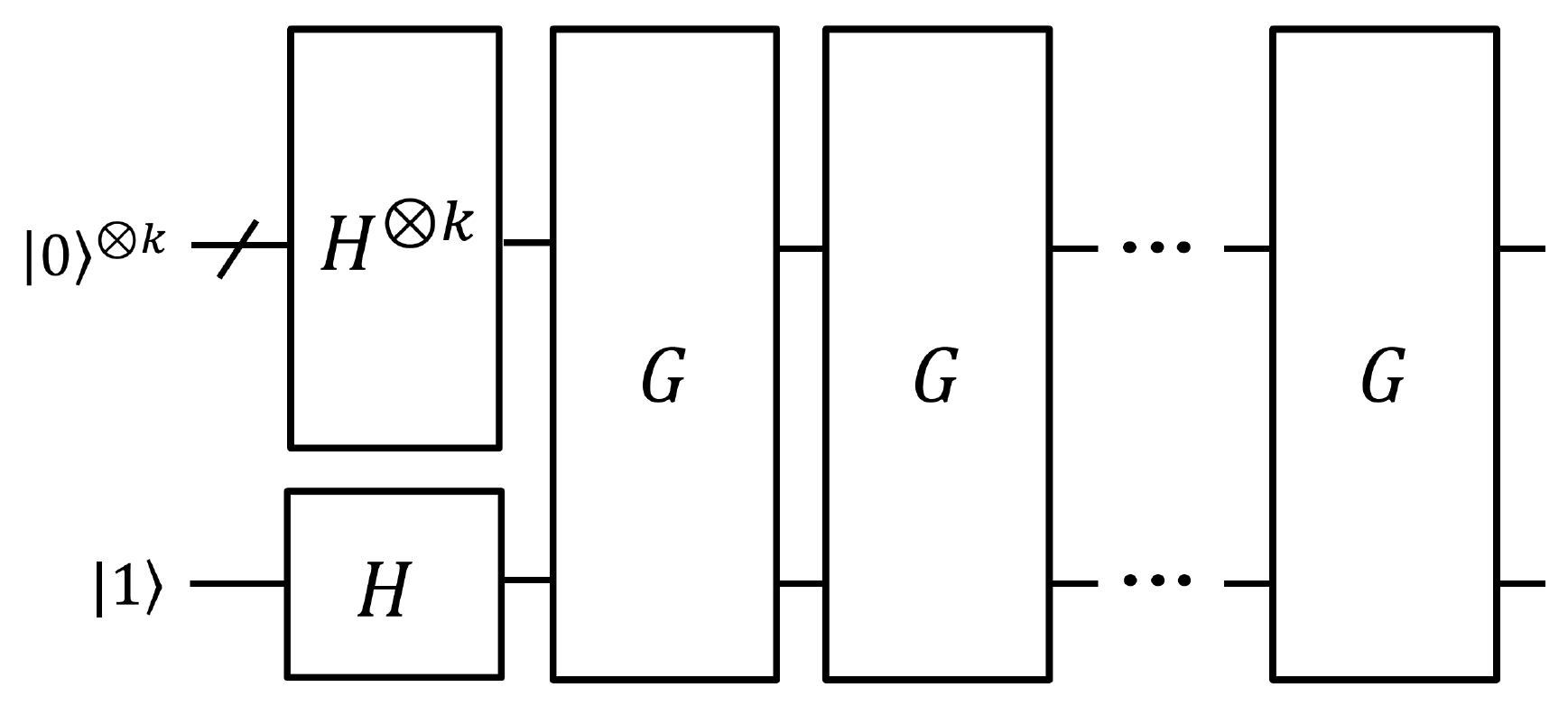} &
\includegraphics[height=1.1in]{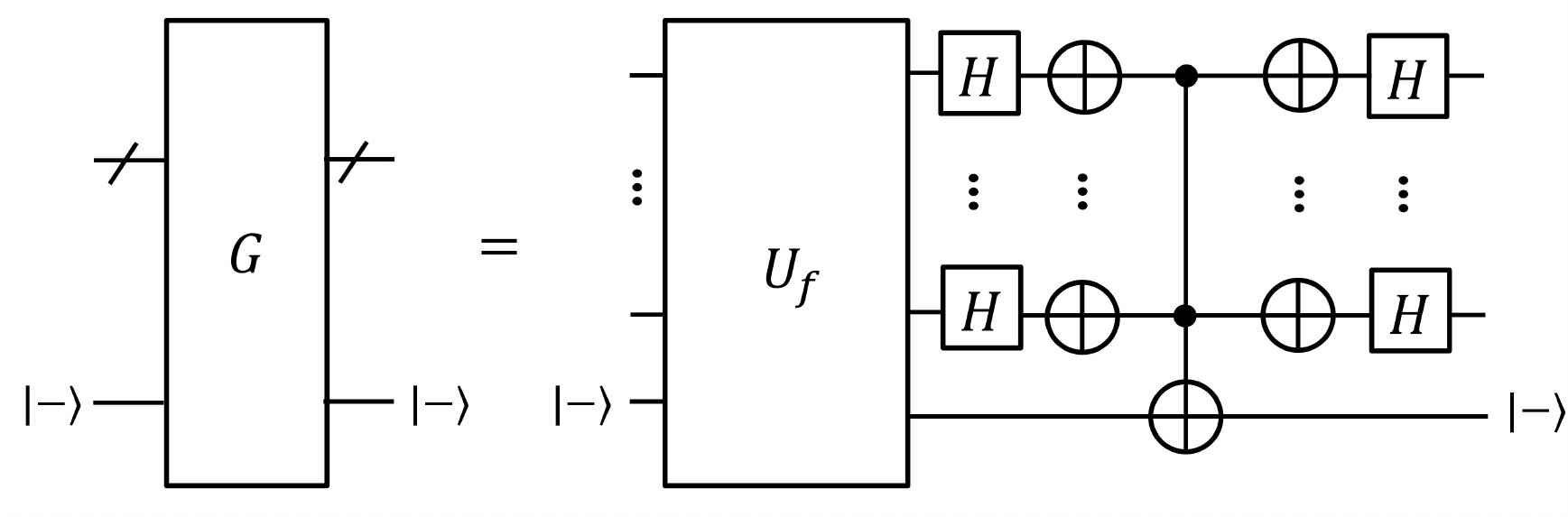}\\
(a) & (b) 
\end{tabular}
}
\caption{\label{fig:grover} (a) Quantum circuit to implement Grover's algorithm. The algorithm consists of creating the equal superposition $\sum_x \ket{x}$ in the upper register which for the case of AES has $k=128, 192, 256$ qubits and a single qubit state $\ket{-} = \ket{0}-\ket{1}$ in the lower register. The operator $G$ is the Grover iterate and is applied a total number of $\lfloor \frac{\pi}{4} \sqrt{2^k} \rfloor$ many times. (b) One round of Grover's algorithm. Shown is the operator $G = U_f \, \left((H^{\otimes k} (2 \ket{0}\bra{0}-{\mathbf 1}_{2^k}) H^{\otimes k}) \otimes {\mathbf 1}_2\right)$ and its circuit decomposition. Note that the effect of the gates between the two layers of Hadamard gates is to invert the phase of the basis state $\ket{0}$ on the upper $k$ bits (up to a global phase).}
\end{figure}

Returning to the case of Grover's algorithm with a unique solution, we
now study the number of gates and the space requirements needed in
order to implement the algorithm. We consider the gates shown in Figure \ref{fig:grover}, in particular we first focus on the circuit shown in part (b) of the figure and analyze its complexity. While $H$ is a Clifford operation, besides the operation $U_f$ which involves the classical computation of (several) AES functions, we also have to
determine the cost $\kappa$ for the operation $(2 \ket{0}\bra{0}-1)$
in equation (\ref{eq:grover}). This reduces to the implementation of
a $k$-fold controlled NOT gate, where for us $k\in\{128,192,256\}$. The resource
estimates for this gates in terms of Toffoli gates can be obtained
from \cite{BBC+:95} to be (as $n\geq 5$): $8k-24$ Toffoli gates which
evaluates to $1,000$, $1,512$, and $2,024$ Toffoli gates per phase operation $(2
\ket{0}\bra{0}-{\mathbf 1}_{2^k})$, respectively. For the number of Clifford$+T$ gates
(counting only $T$s) one could directly apply an upper bound by
multiplying $\kappa$ with $7$, however, one can derive a slightly
better bound: as shown in \cite{WR:2014} (see also
\cite{Maslov:2015}), one can employ phase cancellations and show an
upper bound of $32k-84$ for a $k$-fold controlled NOT gate, i.e., we
obtain $4,012$, $6,060$, and $8,108$ for the $T$-count per phase operation for the three key sizes $k\in \{128, 192, 256\}$. 

We spend the rest of the paper to obtain estimates for $f\colon \{0,1\}^{k} \rightarrow \{0,1\}$
which proceeds by first mapping $K \mapsto
(\mathrm{AES}_K(m_1),\dots,\mathrm{AES}_K(m_r))$ and then computing the equality function of the resulting vector with the given ciphertexts $c_1, \ldots, c_r$, where $c_i \in \{0,1\}^{128}$. In other words, we define the value of $f$ on a given input key $K\in \{0,1\}^k$ (where $k \in \{128,192,256\}$) as follows:  
\[ 
f(K) := (\mathrm{AES}_K(m_1)=c_1) \wedge \ldots \wedge (\mathrm{AES}_K(m_r)=c_r).
\]
As argued below, it is plausible that $r=3,4,5$ are sufficient for the three standardized AES key sizes. The equality function can be implemented by a multiply controlled NOT gate that has $128 r$ (many controls where $r=3,4,5$) and a single target. Using the above formulas this leads to Toffoli counts of $3,048$, $4,072$, and $5,096$, respectively, as well as $T$-counts of $12,204$, $16,300$, and $20,396$, respectively. We return to the question of providing exact quantum resource estimates for Grover's algorithm in Section \ref{subsec:groverEst} after the implementation details of the ``oracle'' function $U_f$ have been derived in the subsequent sections.

\section{Implementing the Boolean predicate---testing a key} 
An essential component needed in Grover's algorithm is a circuit which on input a candidate key $\ket{K}$ indicates if this key is equal to the secret target key or not. To do so, the idea is to simply encrypt some (fixed) plaintext under the candidate key and compare the result with the (assumed to be known) corresponding ciphertext under the secret target key.

\subsection{Ensuring uniqueness of the solution} 
As AES always operates on 128-bit plaintexts, at least for $192$-bit and $256$-bit keys we have to assume that fixing a single plaintext-ciphertext pair is not sufficient to determine a secret key uniquely. 

Arguing with the strict avalanche criterion \cite{Kon81,DGP92} exactly in the same way as in \cite[Section~2.1]{RoSt15}, we can plausibly assume that for every pair of keys ($K,K')\in\{0,1\}^{k\times k}$ with $K\ne K'$ the condition $$(\mathrm{AES}_K(m_1),\dots,\mathrm{AES}_K(m_r))\ne(\mathrm{AES}_{K'}(m_1),\dots,\mathrm{AES}_{K'}(m_r))$$ holds for some suitable collection of plaintexts $m_1,\dots, m_r$. The reason for this is that, for a fixed plaintext, when flipping a bit in the secret key, then each bit of the corresponding ciphertext should change with probability $1/2$. Hence, for $r$ simultaneous plaintext-ciphertext pairs that are encrypted under two secret keys $K'\ne K$ we expect to get different results with probability about $1-2^{-rn}$, if the plaintexts are pairwise different, where $n$ denotes the length of the message. Hence out of a total of $2^{2k}-2^k$ key pairs $(K,K')$ with $K\ne K'$, about $(2^{2k}-2^k)\cdot2^{-rn}\le 2^{2k-rn}$ keys $K'\ne K$ are expected to give the same encryptions. Hence it seems plausible to estimate that
\begin{equation}
r>\lceil 2k/n\rceil\label{equ:kpunicity}
\end{equation}
plaintexts suffice to ensure that for every $K'\ne K$ at least one separating plaintext is available. As AES has $128$-bit plaintexts we have that $n=128$, i.e., Eq.~\eqref{equ:kpunicity} implies that for key length $k$ the adversary has $r>\lceil 2k/128\rceil$ plaintext-ciphertexts pairs $(m_1,r_1),\dots,(m_r,c_r)$ for the target key available. In other words, to characterize the secret target key uniquely, we assume that $r=3$ (AES-$128$), $r=4$ (AES-$192$), and $r=5$ (AES-$256$) suitable plaintext-ciphertext pairs are known by the adversary.

\subsection{Reversible and quantum circuits to implement AES}
 
We assume that the reader is familiar with the basic components of AES. For a detailed specification of AES we refer to FIPS-PUB 197~\cite{FIPS01}. To realize this round-oriented block cipher as a reversible circuit over the Toffoli gate set, respectively as a quantum circuit over the Clifford$+T$ gate set, we need to take care of the \emph{key expansion}, which provides all needed 128-bit round keys, as well as the individual rounds. While the number of rounds depends on the specific key length $k$, the four main functions---{\tt AddRoundKey}, {\tt MixColumns}, {\tt ShiftRows}, and {\tt SubBytes}---that are used to modify the 128-bit internal state of AES are independent of $k$.

First, we discuss the realization of these four functions, before going into details of combining them with the key expansion into complete round functions and a full AES. In our design choices, we tried to keep the number of qubits low, even when this results in a somewhat larger gate complexity. For instance, to implement the ${\mathbb F}_{256}$-multiplications within {\tt SubBytes}, we opted for a multiplier architecture requiring less qubits, but more Clifford and more $T$-gates.

\subsubsection{Circuits for the basic AES operations} 
The internal AES state consists of 128~bits, organized into a rectangular array of $4\times 4$~bytes. We will devote 128~qubits to hold the current internal state.

\paragraph{{\tt AddRoundKey}.} In the implementation of the key expansion, we ensure that the current round key is available on 128 dedicated wires. Implementing the bit-wise XOR of the round key then reduces to 128~CNOT gates which can all be executed in parallel.

\paragraph{{\tt MixColumns}.} Since {\tt MixColumns} operates on an entire column of the state or 32~(qu)bits at a time, the matrix specified in~\cite{FIPS01} was used to generate a $32\times 32$ matrix. 
An LUP-type decomposition was used on this $32\times 32$~matrix in order to compute this operation in place with 277 CNOT gates and a total depth of 39.
\ Example~\ref{ex:exam1} offers a similar but smaller version of an LUP-type decomposition as we used.

\paragraph{{\tt ShiftRows}.}  As {\tt ShiftRows} amount to a particular permutation of the current AES state, we do not have to add any gates to implement this operation as it corresponds to a permutation of the qubits. Instead, we simply adjust the position of subsequent gates to make sure that the correct input wire is used.

\paragraph{{\tt SubBytes}.} This operation replaces one byte of the current state with a new value. For a classical implementation, a look-up table can be an attractive implementation option, but for our purposes, explicitly calculating the result of this operation seems the more resource friendly option. Treating a state byte as element $\alpha\in {\mathbb F}_{2}[x]/(1+x+x^3 + x^4 + x^8)$, first the multiplicative inverse of $\alpha$ (leaving $0$ invariant) needs to be found. This is followed by an affine transformation. To find $\alpha^{-1}$ we adopt the idea of \cite{ARS13} to build on a classical Itoh-Tsujii multiplier, but we work with in-place matrix multiplications. Specifically, we compute
\begin{equation}
\alpha^{-1}=\alpha^{254}=((\alpha \cdot \alpha^2)\cdot  (\alpha \cdot \alpha^2)^4 \cdot  (\alpha \cdot \alpha^2)^{16}\cdot \alpha^{64})^2,\label{equ:howtoinvert}
\end{equation}
exploiting that all occurring exponentiations are ${\mathbb F}_2$-linear. Using again an LUP-type decomposition, the corresponding matrix-multiplication can be realized in-place, using CNOT gates only. And by adjusting the positions of subsequent gates accordingly, realizing the permutation is for free, no gates need to be introduced for this.

\begin{figure}[hbt]
\centerline{\includegraphics[width=0.7\textwidth]{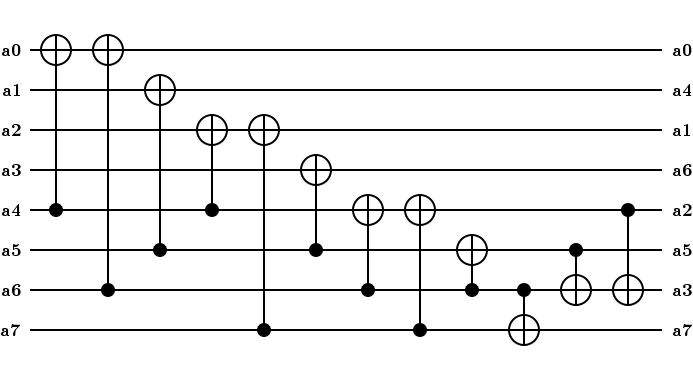}}
\caption{Squaring in ${\mathbb F}_2[x]/(1+x+x^3+x^4+x^8)$}\label{fig:squaring}
\end{figure}

\begin{example}\label{ex:exam1} Squaring in ${\mathbb F}_{2}[x]/(1+x+x^3 + x^4 + x^8)$ can be expressed as multiplying the coefficient vector from the left with

$$\begin{bmatrix}
  1 & 0 & 0 & 0 & 1 & 0 & 1 & 0 \\
  0 & 0 & 0 & 0 & 1 & 0 & 1 & 1 \\
  0 & 1 & 0 & 0 & 0 & 1 & 0 & 0 \\
  0 & 0 & 0 & 0 & 1 & 1 & 1 & 1 \\
  0 & 0 & 1 & 0 & 1 & 0 & 0 & 1 \\
  0 & 0 & 0 & 0 & 0 & 1 & 1 & 0 \\
  0 & 0 & 0 & 1 & 0 & 1 & 0 & 0 \\
  0 & 0 & 0 & 0 & 0 & 0 & 1 & 1 \\
 \end{bmatrix}=\begin{bmatrix}
  1 & 0 & 0 & 0 & 0 & 0 & 0 & 0 \\
  0 & 0 & 0 & 0 & 1 & 0 & 0 & 0 \\
  0 & 1 & 0 & 0 & 0 & 0 & 0 & 0 \\
  0 & 0 & 0 & 0 & 0 & 0 & 1 & 0 \\
  0 & 0 & 1 & 0 & 0 & 0 & 0 & 0 \\
  0 & 0 & 0 & 0 & 0 & 1 & 0 & 0 \\
  0 & 0 & 0 & 1 & 0 & 0 & 0 & 0 \\
  0 & 0 & 0 & 0 & 0 & 0 & 0 & 1 \\
 \end{bmatrix}
\cdot\begin{bmatrix}
  1 & 0 & 0 & 0 & 0 & 0 & 0 & 0 \\
  0 & 1 & 0 & 0 & 0 & 0 & 0 & 0 \\
  0 & 0 & 1 & 0 & 0 & 0 & 0 & 0 \\
  0 & 0 & 0 & 1 & 0 & 0 & 0 & 0 \\
  0 & 0 & 0 & 0 & 1 & 0 & 0 & 0 \\
  0 & 0 & 0 & 0 & 0 & 1 & 0 & 0 \\
  0 & 0 & 0 & 0 & 1 & 1 & 1 & 0 \\
  0 & 0 & 0 & 0 & 0 & 0 & 1 & 1 \\
 \end{bmatrix}
\cdot\begin{bmatrix}
  1 & 0 & 0 & 0 & 1 & 0 & 1 & 0 \\
  0 & 1 & 0 & 0 & 0 & 1 & 0 & 0 \\
  0 & 0 & 1 & 0 & 1 & 0 & 0 & 1 \\
  0 & 0 & 0 & 1 & 0 & 1 & 0 & 0 \\
  0 & 0 & 0 & 0 & 1 & 0 & 1 & 1 \\
  0 & 0 & 0 & 0 & 0 & 1 & 1 & 0 \\
  0 & 0 & 0 & 0 & 0 & 0 & 1 & 0 \\
  0 & 0 & 0 & 0 & 0 & 0 & 0 & 1 \\
 \end{bmatrix}\quad.$$
From this, we see that in-place-squaring can be implemented with only twelve CNOT gates. The resulting circuit is shown in Figure~\ref{fig:squaring}.
\end{example}

To realize the six multiplications in Equation~\eqref{equ:howtoinvert}, we use a general purpose multiplier in the underlying binary field. We opted for a design by Maslov et al. \cite{MMCP09}, which requires less than 60\% of the number of qubits than a more recent design in \cite{KeSt15}. This comes at the cost of an increased gate complexity, however, and a different design choice could be considered. For the specific polynomial basis representation of ${\mathbb F}_{256}$ at hand, Maslov et al.'s design, requires $64$~Toffoli plus $21$~CNOT gates, which with Amy et al. \cite{AMMR13} translates into $64\cdot 7=448$~$T$- plus $64\cdot 8+21=533$~Clifford gates.

Noticing that three of the multiplications in  Equation~\eqref{equ:howtoinvert} are actually duplicates, it turns out that four multiplications suffice in order to implement the inversion. Trying to reduce the number of total qubits required at each step, the actual calculation of computing $\alpha^{-1}$ fits into 40~qubits total, producing $\ket{\alpha}$, $\ket\alpha^{-1}$, and twenty-four reinitialized qubits as output. To do so, and reinitialize qubits, we invest twelve linear transformations and eight ${\mathbb F}_{256}$-multiplications, totalling $3584$~$T$-gates and $4539$ Clifford gates.

Once $\alpha^{-1}$ is found, the affine transformation specified in~\cite{FIPS01} must be computed, which can be done with an LUP-type decomposition; four uncontrolled NOT gates take care of the vector addition after multiplication with a matrix. In total one $8$-bit S-box requires $3584$~$T$-gates and $4569$~Clifford gates.

\paragraph{{\tt SubBytes}---an alternative implementation minimizing
  qubits.} The inversion $\alpha\mapsto \alpha^{-1}$ (where $0$ is
mapped to $0$) can be seen as a permutation on ${\mathbb
  F}_{256}$. This permutation is odd, while quantum circuits with NOT,
CNOT, and Toffoli gates on $n>3$ qubits generate the full alternating
group $A_{2^n}$ of even permutations.  Hence we have to use one
ancilla qubit, i.e., nine qubits in total.  The task is then to
express a permutation on $512$ points in terms of the generators
corresponding to the NOT, CNOT, and Toffoli gates.  While computer
algebra systems like Magma \cite{WCP97} have built-in functions for
this, the resulting expressions will be huge.  In order to find a
short factorization, we compute a stabilizer chain and corresponding
transversals using techniques similar to those described in
\cite{EP98}.  We use a randomized search to find short elements in
each transversal.  As it is only relevant to implement the exact
function when the ancilla qubit is in the state $|0\rangle$, we choose
the first $256$ points in the basis for the permutation group as those
with the ancilla in the state $|0\rangle$, and the remaining $256$
points as those with the ancilla in the state $|1\rangle$.  This
allows to compute a factorization modulo permutations of the last
$256$ points.  With this approach, we found a circuit with no more
than $9695$ $T$-gates and $12631$ Clifford gates, less than three
times more gates than the version above, but using only $9$ instead of
$40$ qubits in total.

\subsubsection{Key Expansion} Standard implementation of the key expansion for AES-$k$ ($k=128, 192, 256$) separates the original $k$-bit key into 4, 6 or 8 \emph{words} of length 32, respectively and must expand the $k$-bit key into forty-four \emph{words} for $k=128$, fifty-two \emph{words} for $k=192$ and sixty \emph{words} for $k=256$. Each AES key expansion uses the same operations and there are only slight differences in the actual round key construction. The operations are {\tt RotWord}, a simple rotation, {\tt SubBytes}, and {\tt Rcon}[$i$], which adds $x^{i-1}\in{\mathbb F}_{256}$ to the first byte of each word.

While the three different versions of AES employ up to 14 rounds of computation, the key expansion is independent of the input. The  \emph{words} created by the key expansion were divided into two categories: the \emph{words} needing {\tt SubBytes} in their computation and those that do not.  The \emph{words} not involving {\tt SubBytes} can be recursively constructed from those that do by a combination of XORings making them simple to compute as needed, saving up to 75\% of the storage cost of the key expansion. The most expensive of these is \emph{word 41} or $w_{41}$ in AES-128 which is constructed by XORing 11 previous \emph{words} costing 352 CNOT gates and a total depth of 11.

\begin{table}[hbt]
\normalsize
\centerline{
\begin{tabular}{ c @{\quad} c @{\quad} c @{\quad} c @{\quad} c @{\quad} c @{\quad} c @{\quad} c }
 \hline\hline
 & \multicolumn{3}{c}{\hspace*{-0.8cm}$\#$gates} & \multicolumn{2}{c}{\hspace*{-0.2cm}depth} & \multicolumn{2}{c}{$\#$qubits} \\
 & NOT & CNOT & Toffoli & $T$ & overall & storage & ancillae \\
 \hline\\[-2ex]
128 & 176 & 21,448 & 20,480 & 5,760 & 12,636 & 320 & 96 \\
192 & 136 & 17,568 & 16,384 & 4,608 & 10,107 & 256 & 96 \\
256 & 215 & 27,492 & 26,624 & 7,488 &  16,408 & 416 & 96 \\[0.2ex]
\hline
\hline
\end{tabular}\bsp
}
\caption{\label{fig:aesKeygen} Quantum resource estimates for the key expansion phase of AES-$k$, where $k \in \{128,192,256\}$.} 
\end{table}\vspace*{-3.5Ex}

Since {\tt SubBytes} is costly, the remaining \emph{words} are stored as they are constructed. In a classical AES implementation, these \emph{words} (every fourth or sixth) are produced by starting with the previous word, however in this construction the previous word must be constructed, and removed, as needed. For example, in AES-128, to construct $w_{8}$, first $w_{7}$ must be constructed as follows: $
w_{7} = w_{4}\oplus w_{3} \oplus w_{2} \oplus w_{1}.
$

This can be done on the previously constructed word (here $w_{4}$) saving qubits, gates, and depth. Since the construction of $w_{8}$ involves the use of $w_{4}$ the above process needs to be repeated to be removed before the end of construction of $w_{8}$. For the construction of these words, similar to {\tt ShiftRows}, {\tt RotWord} can be eliminated if the position of the gates is shifted to use the correct wires. Since {\tt SubWord} applies {\tt SubBytes} to each byte of the word independently, each of the four {\tt SubBytes} computations can be done concurrently. 

\begin{example} Below is the construction of $w_8$. Notice that $w_7$ is constructed on top of $w_4$.
\[
\centerline{\includegraphics[width=0.95\textwidth]{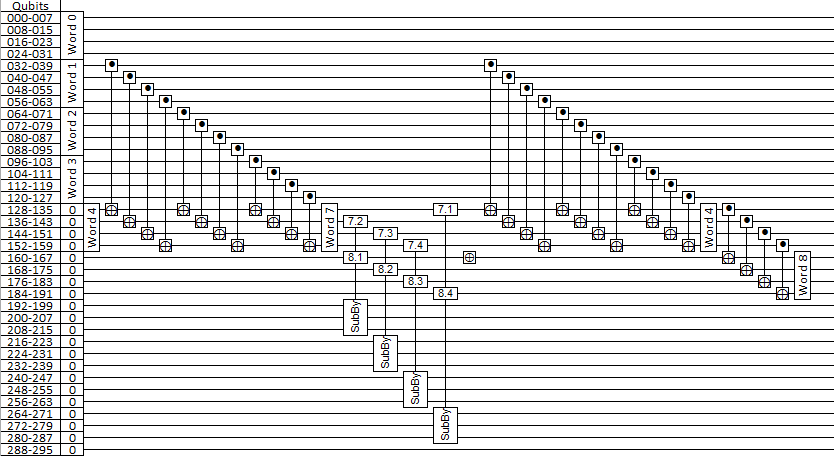}}
\]

\end{example}

To allow each of the four {\tt SubBytes} routines per round to perform simultaneously, 96 auxilary qubits would be needed, along with the 32 needed to store the new word. With each word constructed requiring the previous word be constructed first, we did not reduce the depth further. Computation costs are listed in Table \ref{fig:aesKeygen} (the listed qubit costs do not include storing the original key). 


\vspace*{-1.8Ex}
\subsubsection{AES Rounds.} 
AES starts with a simple whitening step---XORing the input with the first four words of the key. Since, in this case, the input is a fixed value, and adding a fixed value can be done by simply flipping bits, approximately 64 uncontrolled NOT gates are used on the first four key words to start round one. This can be reversed later when needed, but saves 128 qubits. If this is not the case, then 128 qubits are needed to store the input and 128 CNOT gates can be used to compute this step. While the 10, 12, or 14 rounds of AES all apply the same basic functions, the circuit structure differs slightly per round to reduce qubits and depth.  {\tt SubBytes} must be computed 16 times per round, requiring 384 auxiliary qubits for all to be done simultaneously or an increase in depth is needed. Using only the minimum 24 auxiliary qubits and the 128 qubits needed to store the result, it was noticed that all 16 {\tt SubBytes} calculations per round could be done with a maximum depth of 8 {\tt SubBytes} cycles.

Since {\tt SubBytes} is not done in place, and AES-$k$ requires 128 qubits per round, the computation takes 128 qubits times the number of rounds per AES, in addition to the number of qubits needed to store the original key. This number can be reduced by reversing steps between computations to clear qubits for future use. Once {\tt SubBytes} has been applied, the input can be removed by reversing enough steps (but the output could not be removed as its counterpart (inverse) is gone). Since AES-128 employs 10 rounds, using 512 qubits for storage and 24 auxiliary qubits, allows the reverse process to be applied three times. For AES-192 and AES-256, we used 640 qubits for storage since we did not manage to have three rounds of reversing on 536 qubits.

\begin{example}The reverse process representation for AES-128. Notice this method leaves Round 4, Round 7 and Round 9 with no way to be removed unless the entire process is reversed.
\[
\centerline{\includegraphics[width=0.95\textwidth]{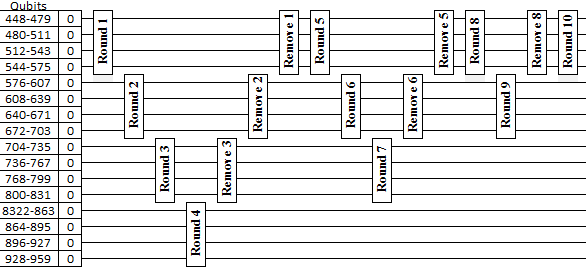}}
\]
 For AES-192 and AES-256 the reversing process is done after rounds five, nine and twelve, requiring only 128 qubits more than AES-128. 
\end{example}

As stated, {\tt ShiftRows} is for free and using an LUP-type decomposition for {\tt MixedColumns} allows this process to be done in place using 277 CNOT gates with a maximum depth of 39. To compute all 10 rounds of AES-128, 536~qubits were needed, 664 qubits were used to compute the 12 rounds of AES-192 and 14 rounds of AES-256.

The XORing of the round keys can be done directly on top of the input for each round. If the round key needed is already constructed, 128 CNOT gates with a depth of 1 are used to complete the round. If the round key is not already constructed and thus a combination of constructed keys, then it only requires this process to be done multiple times. AES-128 requires this to be done 11~times (the most) in the case of $w_{41}$, increasing the depth and CNOT gate count by at most 11.

\subsection{Resource estimates: reversible AES implementation} The numbers listed in the three tables below show the costs in gates, depth and qubits to achieve the output of each AES-$k$ system.\vspace*{-3Ex}

\begin{table}[hbt]
\normalsize
\begin{center}
\begin{tabular}{ c @{\quad} c @{\quad} c @{\quad} c @{\quad} c @{\quad} c  }
 \hline\hline
 & \multicolumn{2}{c}{\hspace*{-0.6cm}$\#$gates} & \multicolumn{2}{c}{\hspace*{-0.3cm}depth} & {$\#$qubits} \\
 & $T$ & Clifford & $T$ & overall &  \\
 \hline\\[-2ex]
Initial & 0 & 0 & 0 & 0 & 128 \\
Key Gen & 143,360 & 185,464  & 5,760 & 12,626 & 320  \\
10 Rounds & 917,504 & 1,194,956 & 44,928  & 98,173 & 536  \\[0.2ex]
 \hline\\[-2ex]
Total & 1,060,864 & 1,380,420 & 50,688 & 110,799 & 984  \\[0.2ex]
\hline
\hline
\end{tabular}\bsp
\caption{\label{fig:aes128} Quantum resource estimates for the implementation of AES-$128$.} 
\end{center}
\end{table}

\begin{table}[hbt]
\normalsize
\begin{center}
\begin{tabular}{ c @{\quad} c @{\quad} c @{\quad} c @{\quad} c @{\quad} c  }
 \hline\hline
 & \multicolumn{2}{c}{\hspace*{-0.6cm}$\#$gates} & \multicolumn{2}{c}{\hspace*{-0.3cm}depth} & {$\#$qubits} \\
 & $T$ & Clifford & $T$ & overall &  \\
 \hline\\[-2ex]
Initial & 0 & 0 & 0 & 0 & 192 \\
Key Gen & 114,688 & 148,776  & 4,608 & 10,107 & 256  \\
12 Rounds & 1,089,536 & 1,418,520 & 39,744 & 86,849 & 664\\[0.2ex]
 \hline\\[-2ex]
Total & 1,204,224 & 1,567,296 & 44,352 & 96,956 & 1,112  \\[0.2ex]
\hline
\hline
\end{tabular}\bsp
\caption{\label{fig:aes192} Quantum resource estimates for the implementation of AES-$192$. The lower gate count in Key Gen and the lower depth, when compared to AES-128, arises from using the additional available space to store intermediate results and to parallelize parts of the circuit.} 
\end{center}
\end{table}
\begin{table}[hbt]
\normalsize
\begin{center}
\begin{tabular}{ c @{\quad} c @{\qquad} c @{\quad} c @{\quad} c @{\quad\quad} c  }
 \hline\hline
 & \multicolumn{2}{c}{\hspace*{-0.6cm}$\#$gates} & \multicolumn{2}{c}{\hspace*{-0.3cm}depth} & {$\#$qubits} \\
 & $T$ & Clifford & $T$ & overall &  \\
 \hline\\[-2ex]
Initial & 0 & 0 & 0 & 0 & 256 \\
Key Gen & 186,368 & 240,699  & 7,488 & 16,408& 416  \\
14 Rounds & 1,318,912 & 1,715,400 & 52,416 & 114,521 & 664\\[0.2ex]
 \hline\\[-2ex]
Total & 1,505,280 & 1,956,099 & 59,904 & 130,929 & 1,336  \\[0.2ex]
\hline
\hline
\end{tabular}\bsp
\caption{\label{fig:aes256} Quantum resource estimates for the implementation of AES-$256$.} 
\end{center}
\end{table}

 
\subsection{Resource estimates: Grover algorithm}\label{subsec:groverEst}

\begin{figure}[hbt]
\centerline{\includegraphics[width=0.7\textwidth]{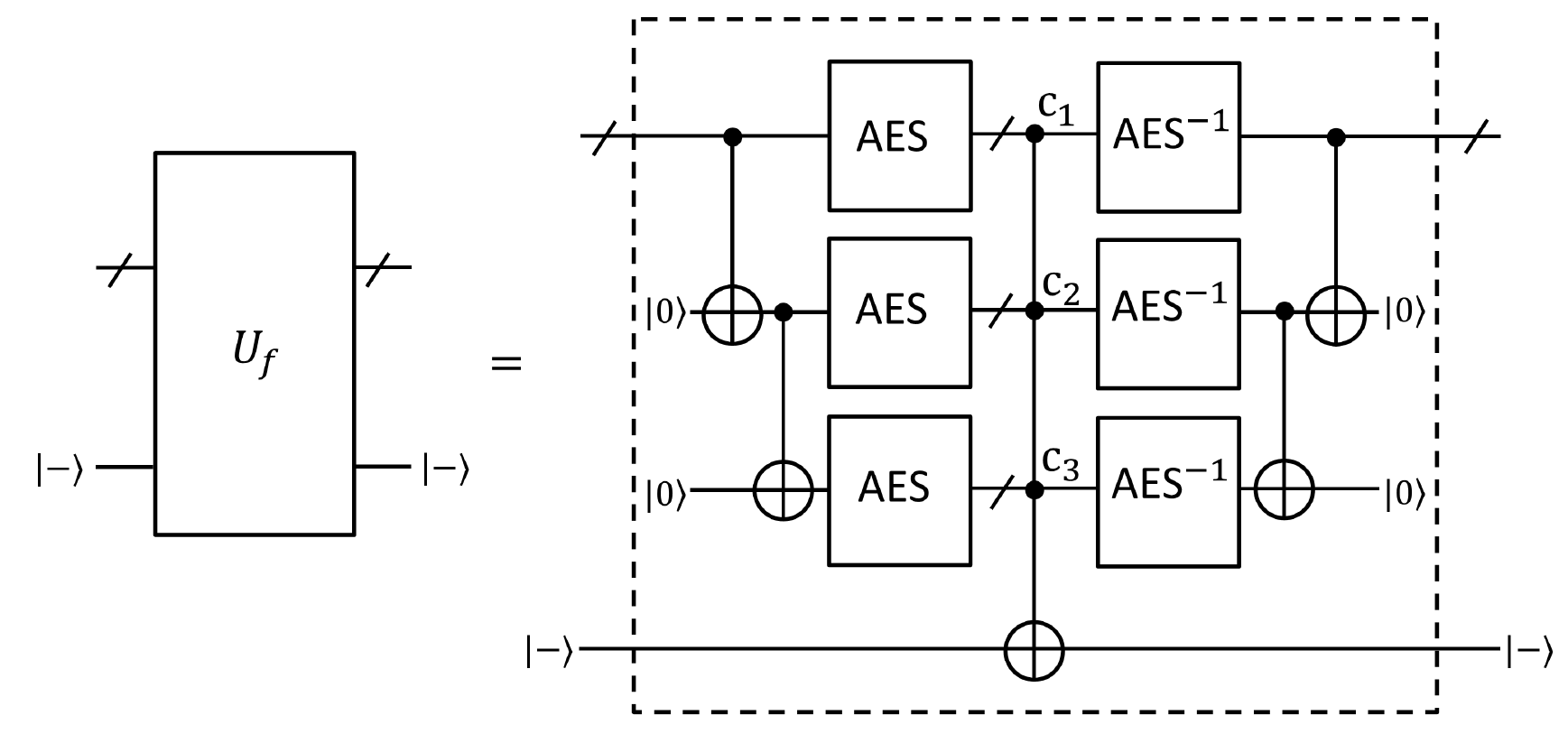}}
\caption{\label{fig:aesBox} The reversible implementation of the function $U_f$ is shown in further detail. In this case the key size $k=128$ is considered for which $r=3$ invocations of AES suffice in order to make the target key unique. For the cases of $k=192$ the number of parallel AES boxes increases to $r=4$ and for $k=256$ to $r=5$, however, the overall structure of the circuit is common to all key sizes.}
\end{figure}

From the discussion in the previous sections we obtain a reversible circuit for computing $\mathrm{AES}_K(m_i)$, i.e., a circuit ${\mathcal C}$ that implements the operation
$\ket{K}\ket{0} \mapsto \ket{K}\ket{\mathrm{AES}_K(m_i)}$. The overall circuit to implement $U_f$ is shown in Figure~\ref{fig:aesBox}. The AES layer can be applied in parallel, however, as the used ancilla qubits have to be returned clean after each round, we have to uncompute each AES box within each round. Hence the depth (and $T$-depth) increases by a factor of $2$ within each invocation of $U_f$. The total number of gates (and $T$-gates) on the other hand increases by a factor of $2r$ as all boxes have now to be counted. The number of qubits is given by $r$ times the number of qubits within each AES box. 

Once the AES boxes have been computed, the result is compared with the given ciphertexts $c_1, \ldots, c_r$. Note that as AES operates on plaintexts/ciphertexts of length $128$ we have that $c_i\in \{0,1\}^{128}$ throughout. The comparison is done by a multiply controlled NOT gate and the controls are either $0$ or $1$ depending on the bits of $c_i$. This is denoted by the superscript $c_i$ on top of the controls in Figure \ref{fig:aesBox}. We can now put everything together to estimate the cost for Grover's algorithm based on the AES-$k$ resource estimates given in the previous section: denoting by $s_k$ the total number of qubits, $t_k$ the total number of
$T$-gates, $c_k$ the total number of Clifford gates, $\delta_k$ the overall $T$-depth and $\Delta_k$ the overall depth, where 
$k=128,192,256$, then we obtain the following estimates for the overall Grover algorithm. The space requirements are $3
s_{128} + 1$ qubits for AES-$128$, $4 s_{192} + 1$ qubits for
AES-$192$, and $5 s_{256} + 1$ qubits for AES-$256$. 

Regarding the
time complexity, we obtain that per Grover iteration we need $6t_{128}$ many $T$-gates for AES-$128$ plus the number of $T$-gates needed for the $384$-fold controlled NOT inside $U_f$ and the $128$-fold controlled NOT to implement the phase $(2\ket{0}\bra{0}-1)$. We estimated the $T$-counts of these two operations earlier to be 12,204 and 1,000 respectively. Overall, we have to perform
$\lfloor \frac{\pi}{4} \; 2^{k/2} \rfloor$ iterations, i.e., we obtain
for the overall $T$-gate count for Grover on AES-$128$ the estimate of
\[
\left\lfloor \frac{\pi}{4} \; 2^{64} \right\rfloor \cdot \big(6 t_{128} + 13,204\big) = 9.24 \cdot 10^{25} = 1.19 \cdot 2^{86}
\]
many $T$-gates. Similarly, we can estimate the number of Clifford gates which for simplicity we just assume to be $6 c_{128}$, ignoring some of the Clifford gates used during the rounds. For AES-$192$ we have to perform $\lfloor \frac{\pi}{4} \; 2^{96} \rfloor$ iterations and for AES-$256$ we have to perform 
$\lfloor \frac{\pi}{4} \; 2^{128} \rfloor$ iterations. For the $T$-count of the controlled operations we obtained $16,300+1,512=17,812$ and $20,396+2024=22,420$ earlier. Overall, this gives for Grover on AES-$192$ the estimate of $3.75 \cdot 10^{36} = 1.81 \cdot 2^{114}$ many $T$-gates and for Grover on AES-$256$ the estimate of $4.03 \cdot 10^{45} = 1.41 \cdot 2^{151}$ many $T$-gates. 
For the overall circuit depth we obtain the number of rounds times $2$ times $\delta_k$, respectively $\Delta_k$, ignoring some of the gates which do not contribute significantly to the bottom line. The overall quantum resource estimates are given in Table \ref{fig:groverSummary}. 

\begin{table}[bht]
\normalsize
\begin{center}
\begin{tabular}{ c @{\quad} c @{\quad} c @{\quad} c @{\quad} c @{\quad} c  }
 \hline\hline
 & \multicolumn{2}{c}{\hspace*{-0.6cm}$\#$gates} & \multicolumn{2}{c}{\hspace*{-0.3cm}depth} & {$\#$qubits} \\
$k$ & $T$ & Clifford & $T$ & overall &  \\
 \hline\\[-2ex]
$128$ & $1.19 \cdot 2^{86}$  & $1.55 \cdot 2^{86}$ & $1.06 \cdot 2^{80}$ & $1.16 \cdot 2^{81}$ & $2,953$ \\
$192$ & $1.81 \cdot 2^{118}$ & $1.17 \cdot 2^{119}$ & $1.21 \cdot 2^{112}$ & $1.33 \cdot 2^{113}$ & $4,449$ \\
$256$ & $1.41 \cdot 2^{151}$ & $1.83 \cdot 2^{151}$ & $1.44 \cdot 2^{144}$ & $1.57 \cdot 2^{145}$ & 6,681 \\[0.2ex]
\hline
\hline
\end{tabular}\bsp
\caption{\label{fig:groverSummary} Quantum resource estimates for Grover's algorithm to attack AES-$k$, where $k\in \{128, 192, 256\}$.} 
\end{center}
\end{table}

\section{Conclusion} When realizing AES, only {\tt SubBytes} involves $T$-gates. Moreover, {\tt SubBytes} is called a minimum of 296 times as in AES-128 and up to 420 times in AES-256. As shown above, for all three standardized key lengths, this results in quantum circuits of quite moderate complexity. So it seems prudent to move away from 128-bit keys when expecting the availability of at least a moderate size quantum computer.

As mentioned in the context of the discussion about Grover's algorithm in the presence of an unknown number of solutions, the implementation of the algorithms in \cite{GHT:98} for quantum counting, \cite{BHM+:2002} for general amplitude amplifications, and \cite{YLC:2014} for fixed-point quantum search might lead to space-time tradeoff implementations of the function $f$. This might in particular be beneficial for the circuit mentioned in \cite{YLC:2014} as this does not incur a space overhead and can deal with an unknown number of solutions, provided an upper bound on the number of solutions is known a priori. We leave the question of providing quantum resource estimations for attacking AES and other block ciphers by means of such fixed-point versions of Grover's algorithm for future work. Also an interesting area of future research is the resource cost estimation of recently proposed quantum linear and differential cryptanalysis \cite{KLL+:2015}.  

\section*{Acknowledgments}
BL and RS were supported by AFRL/RIKF Award No. FA8750-15-2-0047. RS was also supported by NATO's Public Diplomacy Division in the framework of ``Science for Peace,'' Project MD.SFPP 984520. The authors thank Schloss Dagstuhl for hosting Seminar 15371, during which part of this work was done.


\end{document}